\documentstyle[12pt,blois,psfig]{article}

\newcommand{\hmsun}{\mbox{$h^{-1}$ $M_{\odot}$}}

\newcommand{\kms}{\mbox{km s$^{-1}$}}

\newcommand{\msun}{\mbox{$M_{\odot}$}}

\newcommand{\kmsmpc}{\mbox{km s$^{-1}$ Mpc$^{-1}$}}

\def\la{\mathrel{\hbox{\rlap{\hbox{\lower4pt\hbox{$\sim$}}}\hbox{$<$}}}}
\def\ga{\mathrel{\hbox{\rlap{\hbox{\lower4pt\hbox{$\sim$}}}\hbox{$>$}}}}

\begin{document}

\heading{The Star Formation History in a Hierarchical Universe}

\author{Rachel S. Somerville $^1$, Joel R. Primack $^2$}
{$^1$ Racah Institute of Physics, The Hebrew University,
Jerusalem 91904, Israel}
{$^2$ Physics Department, University of California, Santa
Cruz, CA 95060 USA}

\begin{bloisabstract}
Observations now probe the star formation history of the Universe back to a
redshift of $z\sim5$. We investigate whether the predictions of semi-analytic
models of galaxy formation based on hierarchical Cold Dark Matter (CDM) type
models are in agreement with these direct observations and also with the
``fossil'' evidence contained in constraints on the ages of present day
early-type galaxies. Previous models \cite{bcfl} predicted that the star
formation rate density falls off rather steeply at $z\ga 2$, and
correspondingly that the majority of the stars in the Universe formed at
relatively low redshift. We investigate the effect of including a bursting mode
of star formation, assuming that galaxy-galaxy mergers trigger starbursts and
using the merger rate that arises naturally in the CDM merging hierarchy. The
resulting starbursts substantially increase the global star formation rate at
high redshift, leading to predictions that are in good agreement with the star
formation rate density at $z\sim3$ obtained from sub-millimeter observations
(SCUBA) and optical/UV estimates after correction for dust extinction. The mass
of stars formed at $z \ge 3$ is correspondingly in better agreement with the
fossil evidence. We also investigate complementary global quantities such as
the mass of cold gas and the average metallicity of cold gas as a function of
redshift, and the integrated extra-galactic background light. We argue that
these observations, taken together, provide strong constraints on the star
formation history of the Universe, and that hierarchical models of the CDM type
are in reasonable agreement with these observations when starbursts are
included.
\end{bloisabstract}

\section{Introduction}

The star formation history of the Universe has been observationally probed
using many approaches, both direct and indirect. The star formation rate
density has been derived from direct observations using a variety of tracers
such as H$\alpha$ \cite{gallego,tresse,kiss}, [OII] \cite{hogg}, or UV emission
\cite{lilly,madau:96,treyer}. Recently, sub-mm observations obtained with the
SCUBA instrument were used to estimate the star formation rate at $z\sim3$
\cite{scuba}. The compilation of these observations, which extend from $z=0$ to
$z\sim5.5$, sketches out an enticing picture of the history of star formation
and have come to be referred to generically as ``the Madau diagram''
\cite{madau:96}.  However, the interpretation of the Madau diagram is not
necessarily straightforward. Because of dust extinction, the optical and UV
tracers certainly underestimate the true star formation rate, but by an
uncertain factor.  The sub-mm flux observed by SCUBA is due to the re-radiation
of light that was absorbed by dust at shorter wavelengths, and so provides a
useful complement to the optical and UV observations. However, the low
resolution of the SCUBA instrument makes it difficult to find unique optical
counterparts and measure redshifts to the sources. The measurement of the mass
density in the form of cold gas as a function of redshift, derived from
observations of damped Lyman-$\alpha$ systems
\cite{lanzetta,storrie-lombardi:96}, provides an indirect and complementary
indicator of the star formation history \cite{fall:96}, as does the measurement
of the average metal content of these systems at various redshifts
\cite{pettini:dlas}. The extragalactic background light (EBL), which has been
measured from the UV to the sub-mm (reviewed by \cite{dwek}), is another
indirect constraint which may prove to be quite powerful, because it
effectively measures the bolometric luminosity density. An alternative method
of determining the star formation history of the Universe, discussed by
A. Renzini at this meeting and elsewhere \cite{renzini:97}, is to use ``fossil
evidence'' such as the age constraints derived from the small scatter in the
observed color-magnitude and fundamental plane relations for early type
galaxies and galactic bulges out to $z\simeq 0.5$ (extensive references are
given in \cite{renzini:97}).
	
A recurring theme of this meeting, and indeed an important question in the
study of galaxy formation today, is whether any existing theoretical framework
can fit these observations into a self-consistent picture. This is a difficult
question, as in order to model these quantities one is forced to consider messy
astrophysical processes such as gas dynamics, star formation, supernovae
feedback, and absorption and re-radiation by dust. Currently, the only
efficient way to do this is using semi-analytic models such as those developed
in e.g. \cite{kwg}, \cite{cafnz}, and \cite{sp98}. Previous work (e.g.,
\cite{cafnz,bcfl}) has left the impression that a ``late epoch of star
formation'' (i.e., the peak in the global star formation history at
$z\sim1.5-2$ and a steep decline at higher redshifts) is a generic feature of
CDM-type models. This behaviour is similar to the star formation history
sketched out in the original Madau diagram \cite{madau:96}, and has been
claimed as a success of these models.

However, there is mounting evidence that the true star formation rate history
at $z\ga3$ may be considerably different from the original estimates. This
could be due to a number of effects, including dust extinction, a steeper slope
on the faint end of the luminosity function (suggested by the results of
\cite{sawicki:97} based on photometric redshifts in the HDF), or unknown
selection effects intrinsic to the Lyman-break selection technique. The revised
star formation rate density may well be consistent with the SCUBA lower limit
at $z=3$. Moreover, as pointed out by Renzini, the integration of the original
Madau diagram, like the models of Ref.~\cite{bcfl} which have a similar star
formation history, would predict that only about 5 percent of the stars in
galaxies today formed before a redshift of three, in fairly serious
disagreement with the fossil evidence (Renzini has argued that $\sim30$ percent
of the stars in galaxies today formed before $z=3$, if cluster galaxies are
typical). This might seem to indicate a serious problem with the entire
hierarchical, CDM-based picture of structure formation.

However, it turns out that a natural feature of the CDM merging hierarchy,
neglected in previous work, can change the situation significantly. This
feature is the star formation occurring in starbursts triggered by
galaxy-galaxy mergers. We first became interested in this idea as a way to
explain the very large luminosities and high star formation rates of the
physically small Lyman-break galaxies \cite{spf98}, which it seems to do very
successfully. In the remainder of this paper, we will summarize the
implications of the starburst model for the properties of Lyman-break galaxies
(LBGs), as well as for the global star formation rate density, cold gas
density, and metal content of the Universe as a function of redshift. We will
conclude that in fact the classic CDM picture is in good agreement with all
these observations within the present uncertainties.

\section{Semi-analytic models of galaxy formation}
Semi-analytic techniques allow one to model the formation and evolution of
galaxies in a hierarchical framework, including the effects of gas cooling,
star formation, supernova feedback, chemical evolution, galaxy-galaxy merging,
and the evolution of stellar populations. The semi-analytic models used here
are described in detail in \cite{mythesis}, \cite{sp98}, and
\cite{spf98}. These models are in reasonably good agreement with a broad range
of local galaxy observations, including the Tully-Fisher relation, the B-band
luminosity function, cold gas contents, metallicities, and colors
\cite{sp98}. Our basic approach is similar in spirit to the models originally
developed by the Munich \cite{kwg} and Durham \cite{cafnz} groups, and
subsequently elaborated by these groups in numerous other papers (reviewed in
\cite{mythesis} and \cite{sp98}). Significant improvements included in
\cite{sp98} are that we assumed a lower stellar mass-to-light ratio (in better
agreement with observed values), included the effects of dust extinction (which
are significant in the B-band), and developed an improved ``disk-halo'' model
for supernovae feedback. With these new ingredients, we simultaneously produce
a Tully-Fisher relation with the correct zero-point and slope, and a B-band
luminosity function with approximately the right normalization and shape on the
bright and faint ends.

The framework of the semi-analytic approach is the ``merging history'' of a
dark matter halo of a given mass, identified at $z=0$ or any other redshift of
interest. We construct Monte-Carlo realizations of halo ``merger trees'' using
the method described in \cite{sk98}. Each branch in the tree represents a halo
merging event. When a halo collapses or merges with a larger halo, we assume
that the associated gas is shock-heated to the virial temperature of the new
halo. This gas then radiates energy and cools. The cooling rate depends on the
density, metallicity, and temperature of the gas. Cold gas is turned into stars
using a simple recipe, and supernova energy reheats the cold gas according to
another recipe. Chemical evolution is traced assuming a constant yield of
metals per unit mass of new stars formed. The spectral energy distribution
(SED) of each galaxy can then obtained by assuming an IMF and using stellar
population models (e.g. \cite{bc93}; here we use the GISSEL98 models
with solar metallicity). It is
then straightforward to obtain magnitudes and colours in any desired filter
band. Our recipes for star formation, feedback, and chemical evolution contain
free parameters, which are set by requiring an average fiducial ``reference
galaxy'' (the central galaxy in a halo with a circular velocity of $220
\,\kms$) to have an I-band magnitude $M_{I} -5\log h = -21.7$ (this requirement
fixes the zero-point of the I-band Tully-Fisher relation to agree with
observations), a cold gas mass $m_{\rm cold} = 10^{10} h^{-2} \msun$, and a
stellar metallicity of about solar. The star formation and feedback processes
are some of the most uncertain elements of these models, and indeed of any
attempt to model galaxy formation. We have investigated several different
combinations of recipes for star formation and supernova feedback (sf/fb) and
also several cosmologies, discussed in detail in \cite{sp98} and \cite{spf98},
but here we will discuss results for a single choice of sf/fb recipe (constant
efficiency quiescent star formation and the ``disk-halo'' feedback model) and
cosmology ($\sigma_8=0.67$ SCDM), and a standard Salpeter IMF.

\begin{figure}
\centerline{\psfig{file=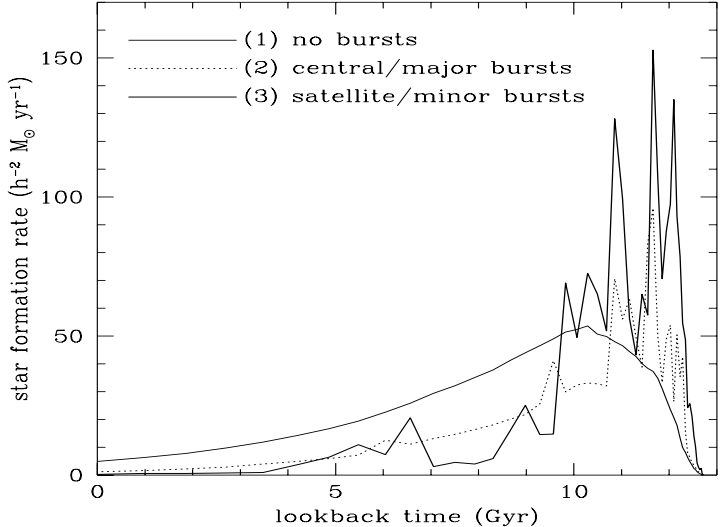,height=7.5truecm,width=10truecm}}
\caption{The total star formation rate for all galaxies that end up within a
typical halo with $V_c = 500\, \kms$ at $z=0$. The solid line shows a model (1)
with no starbursts, the dotted line shows a model (2) with mergers only between
satellite and central galaxies and bursts only in major mergers, and the dashed
line shows a model (3) with satellite-satellite and satellite-central mergers
and bursts in major and minor mergers. The lookback time is computed for
$\Omega=1$ and $H_0=50\,\kmsmpc$. Note that the peaks, representing starbursts,
occur primarily at lookback times of 8 to 12 Gyr (redshifts $z \sim 1-5)$. }
\label{fig:sfhist_group}
\end{figure}

\subsection{Modeling starbursts}
We start with the ansatz that galaxy-galaxy mergers trigger starbursts. This
premise has considerable observational support and is also supported by N-body
simulations with gas dynamics and star formation \cite{mihos}. In our models,
galaxies that are within the same large halo may merge according to two
different processes. Satellite galaxies lose energy and spiral in to the center
of the halo on a {\it dynamical friction} time-scale. In addition, satellite
galaxies orbiting within the same halo may merge with one another according to
a modified {\it mean free path} time-scale. Our modeling of the latter process
is based on the scaling formula derived in \cite{makino-hut} to describe the
results of dissipationless N-body simulations in which galaxy-galaxy encounters
and mergers were simulated, covering a large region of parameter space.

When any two galaxies merge, the ``burst'' mode of star formation is turned on,
with the star formation rate during the burst modeled as a Gaussian function of
time. The burst model has two adjustable parameters, the time-scale of the
burst and the efficiency of the burst (the fraction of the combined cold gas
reservoir of both galaxies that is turned into stars over the duration of the
burst). The timescale and efficiency parameters that we use are based on the
simulations \cite{mihos} mentioned above, treating major ($m_{\rm
smaller}/m_{\rm larger} > f_{\rm bulge} \sim 0.3$) and minor mergers
separately. Details are given in \cite{spf98}. The results of including bursts
in a single halo with $V_c = 500\,\kms$ at $z=0$ are shown in
Fig.~\ref{fig:sfhist_group}.

\subsection{Comoving number density}
\begin{figure}
\centerline{\psfig{file=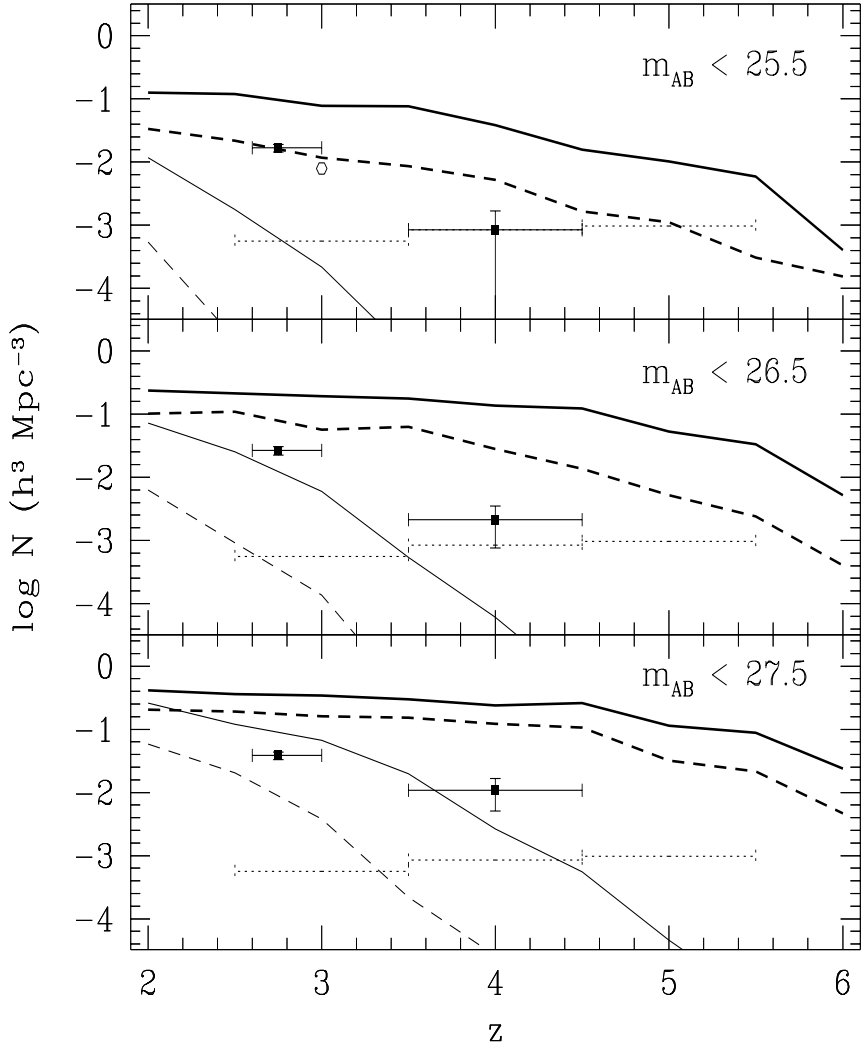,height=15truecm,width=12truecm}}
\caption{The comoving number density of galaxies brighter than $m_{\rm AB}
=25.5$ (top panel), 26.5 (middle panel), and 27.5 (bottom panel), where $m_{\rm
AB} = (V_{606} + I_{814})/2$. The open hexagon indicates the comoving number
density of LBGs from the ground-based sample of Steidel et
al. \protect\cite{adelberger:98}. The filled squares at $z=2.75$ and $z=4$
indicate the number density of U and B drop-outs (respectively) in the HDF
\protect\cite{pozzetti:98}. The dashed horizontal lines correspond to one
galaxy in the HDF volume for the indicated redshift intervals. Bold solid lines
show the comoving number density of galaxies in the SCDM models with
starbursts; light solid lines show the results of the no-burst models. Dashed
lines show the result of reducing the flux of each galaxy by a factor of three
to estimate the effects of dust extinction. In the no-burst models, the
comoving number density of bright LBGs decreases rapidly with increasing
redshift. The burst models predict that the comoving number density at $z=4$ is
nearly as high as at $z=3$. }
\label{fig:counts_scdm}
\end{figure}

\begin{figure}
\centerline{\psfig{file=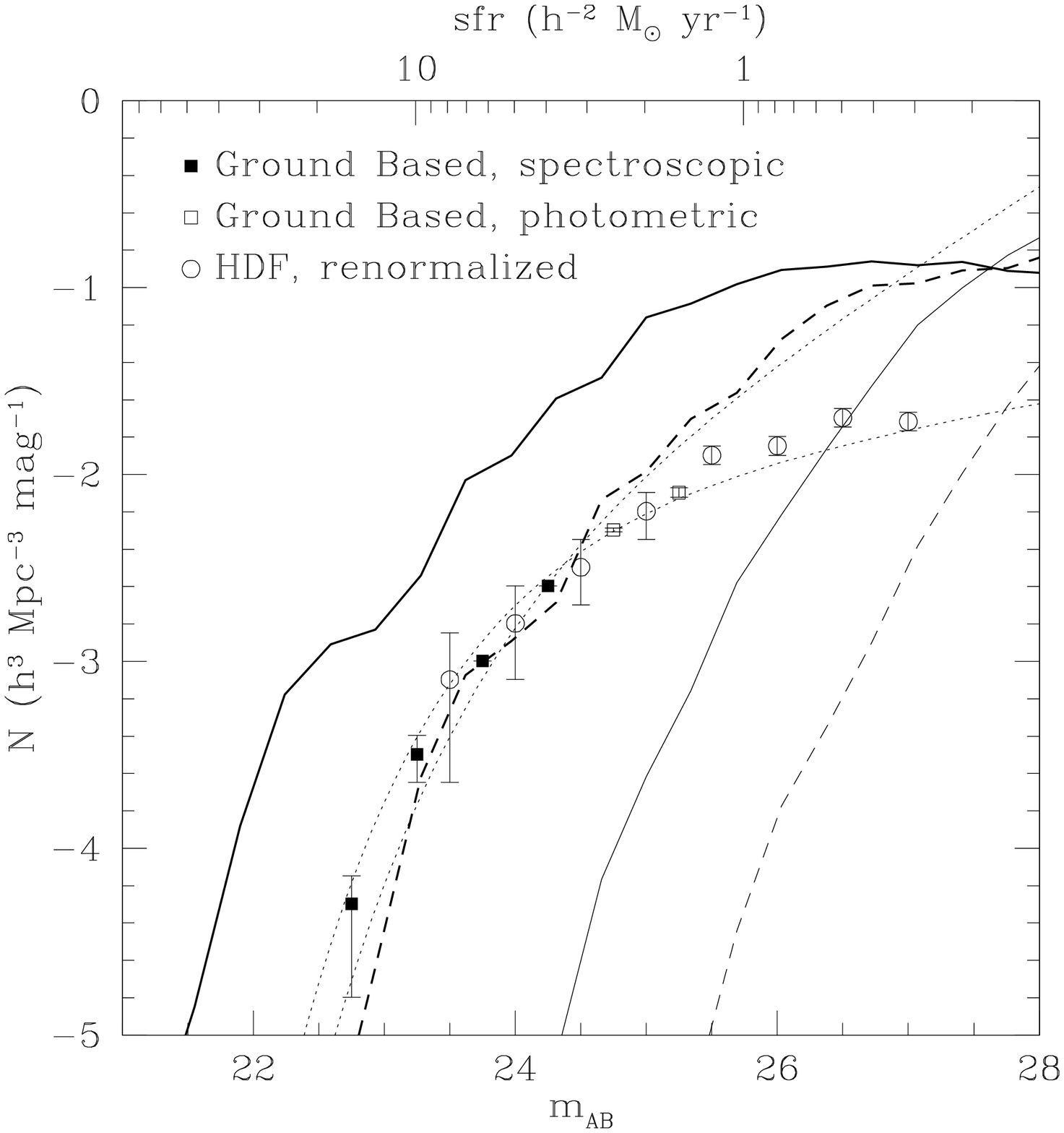,height=8truecm,width=8truecm}}
\caption{The luminosity function at $z=3$ in the SCDM cosmology. Symbols show
the composite luminosity function derived by \protect\cite{dickinson:98} from
the combined ground-based and HDF samples, and the dotted line shows the
Schechter fit to these results ($\alpha=-1.3$). A Schechter function with a
steeper slope, as found by \protect\cite{sawicki:97}, is also shown. Bold lines
indicate the results of the models with starbursts, and light lines indicate
the results of the no-burst models. The dashed lines show the results of a
factor of three correction for dust extinction. }
\label{fig:lf_scdm}
\end{figure}

Fig.~\ref{fig:counts_scdm} shows the comoving number density of galaxies
brighter than a fixed magnitude limit (at rest $\sim 1500$ \AA) as a function
of redshift. Fig.~\ref{fig:counts_scdm} shows that models without bursts
underpredict the number density of bright galaxies, while the burst models
reproduce or exceed the observed number densities. Although tuning the free
parameter that governs the efficiency of quiescent star formation 
can shift these curves up or down, the steep dependence on redshift is
a generic feature of the quiescent models. Similarly, the flatter
dependence arising in the burst models is due to several physical
effects and is not sensitive to the values of the free parameters. The
galaxy-galaxy merger rate is higher at higher redshift, and galaxies
are more gas-rich and produce brighter starbursts. In addition,
following \cite{mihos}, we suppress the starburst mode in
bulge-dominated systems, which become more common at lower redshift.

Internal extinction due to dust is likely to be important, but the amount of
dust and the resulting extinction are quite uncertain. The mean correction
factors at rest 1500 \AA\, estimated by
Refs.~\cite{pettini:dust,dickinson:98,calzetti97b} range from $\sim2$ to
$\sim7$. More dramatic corrections, as large as a factor of $\sim 20$, have
been suggested \cite{meurer,sawickiyee98}. At this meeting, C. Steidel reported
that the net correction currently favored by his group is a factor of $\sim7$
for the ${\cal R}< 25.5$ LBGs, but that the extinction is strongly
differential, i.e. intrinsically brighter galaxies suffer larger
extinctions. Here we have shown a conservative estimate of the effects of dust,
by simply decreasing the luminosity of each galaxy by a factor of three. After
this correction, the no-burst models fail quite dramatically, while the burst
models still appear to be doing rather well.

Fig.~\ref{fig:lf_scdm} shows the rest 1500 \AA\, luminosity function
(equivalently, the star formation rate distribution) at $z=3$. The luminosity
function in the no-burst models is a steep power-law. Increasing the efficiency
of quiescent star formation would move the curve to the left, but the function
is the wrong shape. More realistic, differential dust extinction would only
exacerbate this problem. In the burst models, the shape of the uncorrected
luminosity function is very similar to the one obtained by \cite{dickinson:98}
from a combination of ground based and HDF data. With a factor of three
correction for dust extinction, the burst models are consistent with the
steeper luminosity function ($\alpha=-2.1$) obtained by \cite{sawicki:97} from
the HDF.

Although the burst models are consistent with the observations when we include
the conservative factor of three correction for dust extinction, the larger
corrections and strong differential extinction now favored by the observers
would destroy this good agreement. However, the remaining uncertainties are
large enough that we should not be too concerned about this. If the basic
starburst scenario is correct, the UV light that allows us to identify the LBGs
is produced by young, high mass, and probably low metallicity stars. There are
large uncertainties in the stellar population models in this regime, and the
results are also highly sensitive to the IMF. In addition, although we have
used the best information available to calibrate our treatment of galaxy-galaxy
mergers and starbursts, the modelling should be improved. We 
\cite{kolatt:mergers}
are currently
investigating the frequency and spatial distribution of halo collisions in very
high resolution N-body simulations (based on the method of \cite{kkk}). We are
also extending the studies of gas dynamics and star formation in interacting
galaxies using the same approach as \cite{mihos}. We will cover a larger region
of parameter space, and use galaxy properties more appropriate to the high
redshift systems we are interested in \cite{walker}.

\subsection{Line-widths, ages, and stellar masses}
The velocity dispersions of observed LBGs can be estimated based on the widths
of nebular emission lines such as H$\beta$. Emission lines have been detected
for a few of the brightest LBGs from the ground-based sample. The velocity
dispersions $\sigma$ derived from the observed linewidths are $\sigma=60-80
\,\kms$ for four objects, and $\sim 190\,\kms$ for one object
\cite{pettini:lw}. These values agree well with the burst models, but are in
strong disagreement with the no-burst models, which peak at $\sigma\sim180 \,
\kms$ \cite{spf98}. These measurements may be affected by several biases, which
remain to be unravelled.  Linewidths have been obtained only for the brightest
LBGs, which may have systematically higher internal velocities. This would
increase the discrepancy with no-burst models. The modelling of the velocity
dispersion at the small radii probed by the observations (approximately the
half-light radius) is also uncertain.

The ages and stellar masses of the burst models \cite{spf98} also agree well
with estimates based on multicolor photometry. LBG colors are well fit by young
($<0.1$ Gyr) stellar populations with moderate amounts of dust
\cite{sawickiyee98}.  These young ages imply that stellar masses are also low
($\sim 10^9 M_\odot$). The LBGs in our no-burst models (and Ref.~\cite{bcfl})
are systematically older and more massive than the data indicate.  More
photometry and spectra of LBGs will help to clarify whether models including
starbursts adequately describe the properties of the LBGs.

\begin{figure}
\centerline{\psfig{file=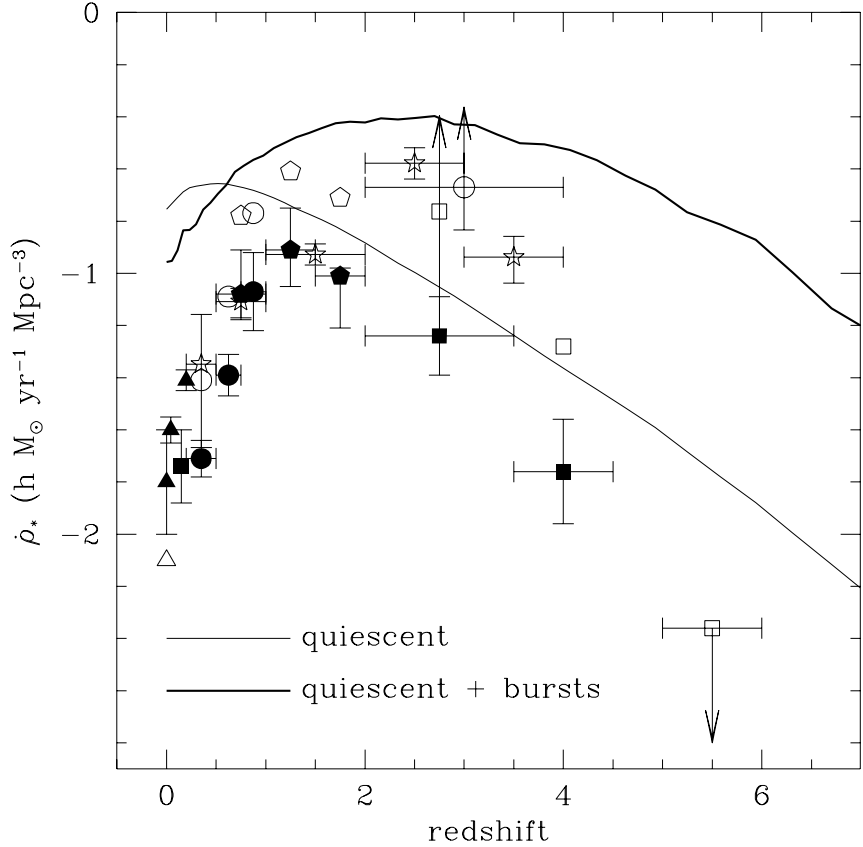,height=12truecm,width=12truecm}}
\caption{The star formation rate density as a function of redshift. The filled
triangle at $z=0$ is from H$\alpha$ observations of local galaxies
\protect\cite{gallego}, with the conversion from H$\alpha$ flux to star
formation rate from Ref.~\protect\cite{madau:98}. The open triangle shows the
star formation rate from the same observations, with the conversion factor
originally used in Ref.~\protect\cite{gallego}. The filled triangle at
$z\sim0.04$ is from new H$\alpha$ observations by the KISS survey
\protect\cite{kiss}, and the filled triangle at $z\sim0.2$ is from H$\alpha$
observations of galaxies from the CFRS \protect\cite{tresse}. The filled square
at $z\sim0.15$ is from 2000 \AA\, fluxes from the UV survey of
\protect\cite{treyer}.  Filled circles are determined from the 2800 \AA\,
fluxes of CFRS galaxies \protect\cite{lilly}, and filled hexagons from the 2800
\AA\, flux of galaxies in the HDF using photometric redshifts
\protect\cite{connolly}. The star symbols are from HDF photometric redshifts
derived by Ref.~\protect\cite{sawicki:97}. Filled squares at $z=2.75$ and $z=4$
are from the $\sim1500$ \AA\, fluxes of U- and B-dropouts in the HDF
\protect\cite{madau:96}. The open symbols show a correction for dust extinction
of a factor of two for 2800 \AA\, flux and three for 1500 \AA\, flux, as
suggested by \protect\cite{pettini:dust}. The tip of the vertical arrow at
$z\sim3$ shows a correction of a factor of seven, which may be indicated by
recent observational studies of dust extinction in LBGs (C. Steidel, this
meeting). The large open circle at $z=3$ shows the lower limit from the SCUBA
observations of the HDF region \cite{scuba} 
(assuming that these sources are stellar rather than AGN).  
Note that the discrepancy between
the predictions and the data at low redshifts ($z\lsim1$) is partly alleviated
when the models are corrected downward by about a factor of 2, to correct for
the overprediction of the number density of $M\lsim 0.1 M_*$ halos in the
Press-Schechter approximation \protect\cite{psover,slkd}. This correction will
also bring the predicted cold gas mass at $z\lsim1$ into better agreement with
the observations (Fig.~6).  }
\label{fig:sfrz}
\end{figure}

\begin{figure}
\centerline{\psfig{file=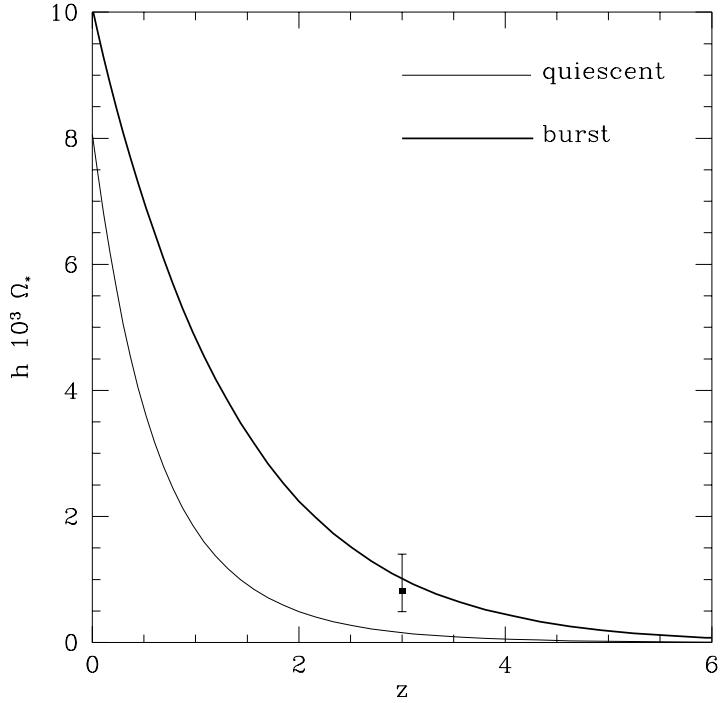,height=10truecm,width=10truecm}}
\caption{The stellar density (in units of the present critical density) as a
function of redshift, in the quiescent (light line) and quiescent + burst (bold
line) models. The data point shows the approximate stellar density at $z=3$
obtained from ``fossil evidence'' (e.g. \cite{renzini:97}).}
\label{fig:omega_star}
\end{figure}
\begin{figure}
\centerline{\psfig{file=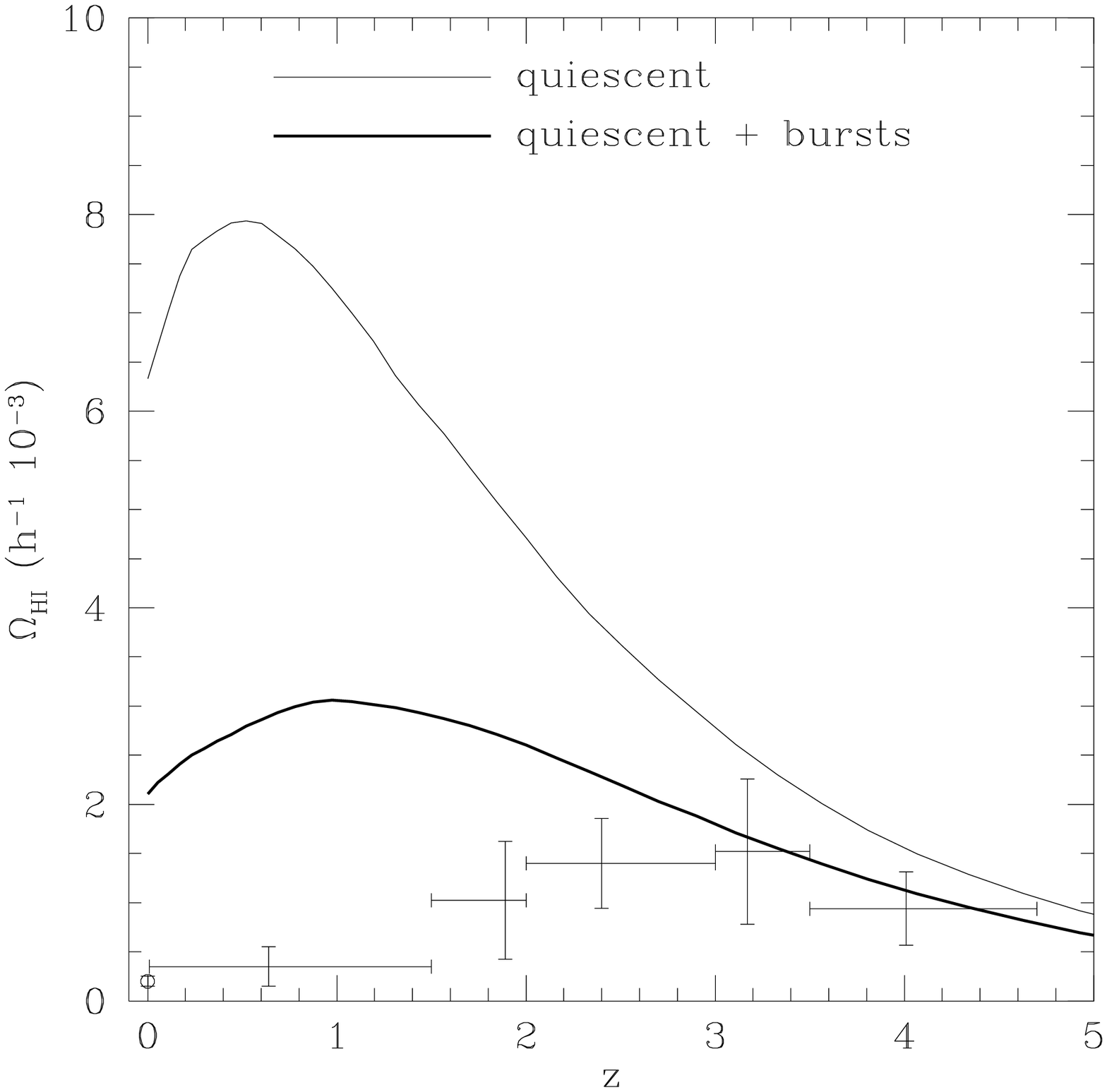,height=10truecm,width=10truecm}}
\caption{The evolution of the mass density of cold gas as a function of
redshift. Data points show the density in the form of H$_{\rm I}$ estimated
from observations of DLAS \protect\cite{storrie-lombardi:96}. The data point at
$z=0$ is from local H$_{\rm I}$ observations \protect\cite{zwaan}. Bold solid
lines correspond to models including bursts, and light lines show the quiescent
models. }
\label{fig:ogasz}
\end{figure}

\subsection{Clustering}
The large ground-based sample of LBGs, with many hundreds of spectroscopic
redshifts, has made it possible to study the clustering of these objects. This
data has provided several indications that the LBGs at $z\sim3$ are strongly
clustered: there are large ``spikes'' in redshift space \cite{steidel:spike},
and the comoving correlation length ($r_0$) obtained from counts in cells is
comparable to that of present day bright galaxies
\cite{giavalisco:98,adelberger:98}. Using dissipationless N-body simulations,
and assuming that one LBG is associated with every halo above a mass threshold
chosen in order to obtain the observed number density of LBGs, the probability
of finding a spike as large as the one reported by \cite{steidel:spike} is then
fairly large for several CDM cosmologies, including both $\Omega=1$ and
low-$\Omega$ variants \cite{wechsler}. The auto-correlation function for these
halos also has a slope ($\gamma$) and correlation length ($r_0$) similar to the
observations \cite{adelberger:98,giavalisco:98}. Similar results have been
obtained by several other groups (e.g. \cite{jingsuto},
\cite{adelberger:98}). It is clear from this work that if the LBGs are
associated with fairly massive halos ($10^{11}-10^{12} \hmsun$, depending on
the cosmology and power spectrum) they will exhibit the observed strong
clustering. The general population of halos with much smaller masses is much
less strongly clustered. The burst models predict that LBGs may be found in
halos with masses as small as $\sim 10^{11} \msun$, and based on the results of
the studies mentioned above, this might appear to be in conflict with the
observations. However, because the starbursts ``turn on'' for a relatively
short period of time ($\sim 50$ Myr), in our picture most of the LBGs are
associated with halos that are merging or have experienced a very recent
merger. When we identify halo collisions in very high resolution N-body
simulations, we find that most 
collisions with enough baryons to produce a
bright starburst involve relatively small mass halos located in dense regions
\cite{kolatt:mergers}. Not surprisingly then, we find that merging or
recently-merged halos are more strongly clustered than the overall population.

\subsection{Star Formation History}
\begin{figure}
\centerline{\psfig{file=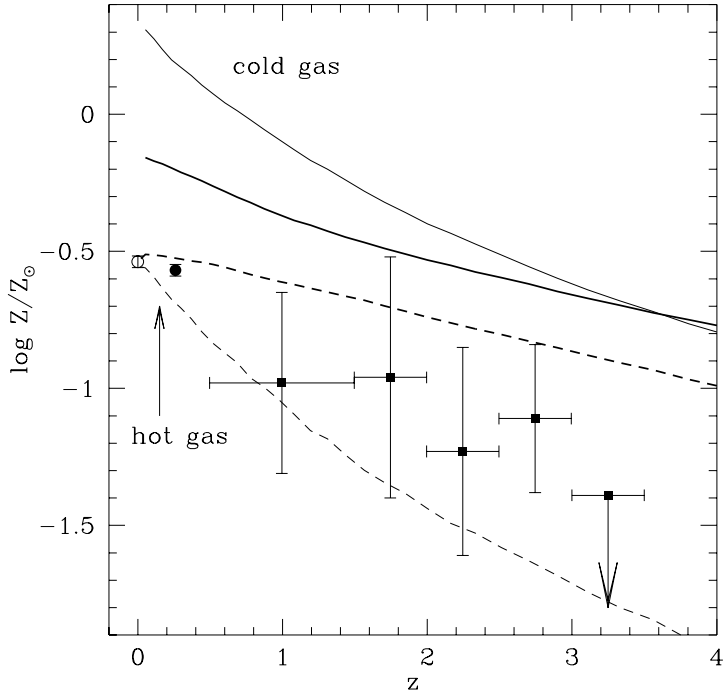,height=10truecm,width=10truecm}}
\caption{Metal content of the Universe. The square symbols are from the
measurements of Zn abundance in DLAS ([Zn/H$_{\rm DLAS}$]) from
\protect\cite{pettini:dlas}. Circular symbols show the metallicity of hot gas
from X-ray observations \protect\cite{mushotsky:97}. Light lines show the
no-burst models and bold lines show the models including starbursts. Solid
lines show the metallicity of cold gas in the models, and dashed lines show the
metallicity of hot gas.}
\label{fig:metz}
\end{figure}
Fig.~\ref{fig:sfrz} summarizes the direct observational estimates of the star
formation rate density of the Universe as a function of redshift, and shows the
model predictions for the quiescent and starburst models.  The models including
only quiescent star formation peak at $z\sim1.5$ and fall off fairly steeply at
higher redshifts, similar to the models of Ref.~\cite{bcfl}. The contribution
of starburst galaxies to the global star formation rate density is clearly
significant at $z\ga2$. The starburst models exhibit a peak at higher redshift
($z\sim3$) and are almost flat out to $z\sim5$. This would appear to be in
better agreement with the UV observations derived from Lyman-break galaxies
after correction for dust extinction, and with the SCUBA results. The poor
behaviour of the models at $z<1$ is partially due to inaccuracies in the
Press-Schechter approximation, and partially a result of our simple assumption
of constant quiescent star formation efficiency.

The stars created at high redshift form the population of old stars in present
day galaxies. Fig.~\ref{fig:omega_star} shows $\Omega_{*}$ as a function of
redshift in the quiescent and burst models. Combining the estimate of the
present day stellar mass density from Ref.~\cite{fukugita:98} with Renzini's
claim that one third of these stars should be in place by $z=3$ leads to the
estimate that h$\Omega_{*}(z=3) = 8 \times 10^{-4}$.  If this estimate is
correct, the no-burst models, like the models of Ref.~\cite{bcfl} or direct
integration of the original Madau diagram \cite{madau:96}, do not produce
enough old stars. The models including starbursts, which are in better
agreement with the ``new Madau diagram'', also produce an old stellar
population that is in good agreement with the ``fossil evidence''.

A complementary quantity is the global mass density in the form of cold
gas. This can be estimated observationally from the frequency and column
densities of damped Lyman-$\alpha$ systems (DLAS). Fig.~\ref{fig:ogasz} shows
the fraction of the critical density in the form of cold gas, as estimated from
DLAS \cite{storrie-lombardi:96} and as predicted in the models. Note that
although the observations only reflect H$_{\rm I}$ gas in systems with column
density greater than $2\times10^{20}$ cm$^{-2}$, the model predictions are for
all cold gas regardless of its form (atomic, ionized, or molecular) or column
density. These observations provide an important constraint on the recipes used
to model quiescent star formation: we cannot simply increase the efficiency of
star formation without limit and still retain enough cold gas to reproduce
these observations. At high redshift, we are already starting to push the
observed lower limit on the amount of cold gas needed to explain the DLAS
observations.

Yet one more indicator of the star formation history is contained in the metal
content of the universe, shown in Fig.~\ref{fig:metz}. The mean metallicity of
cold gas in DLAS at $z\sim1-3$ has been estimated by \cite{pettini:dlas}, and
the metallicity of hot gas in clusters has been estimated from X-ray
observations \cite{mushotsky:97}. Our treatment of chemical evolution is rather
crude, effectively tracing the products of Type II SN only. The effective yield
in the models is adjusted in order to attain an average stellar metallicity of
about solar at $z=0$ in Milky Way sized galaxies, and hot gas with $Z\simeq0.3
Z_{\odot}$ in large halos. The metallicity of the cold gas in the models is not
directly constrained but is determined by the feedback and cooling processes.
Once again the burst models show milder evolution with redshift, which is
directly connected to the flat shape of the star formation history at $z\ga3$
seen in Fig.~\ref{fig:sfrz}. Both the quiescent and burst models are probably
consistent with the DLAS observations, given that the metallicity of the DLAS
may be underestimated by up to a factor of two due to selection effects
(M. Pettini, private communication). Increasing the efficiency of quiescent
star formation would increase the metallicity of the cold gas at high redshift
and violate this constraint, unless a large fraction of the metals are ejected
by supernova winds (unlikely except in very small mass dwarf systems
\cite{maclow_ferrara}). The burst models also predict very little evolution in
the metallicity of hot gas out to $z\sim0.3$, in good agreement with the X-ray
observations.

\subsection{The Extra-galactic Background Light}
\begin{figure}
\centerline{\psfig{file=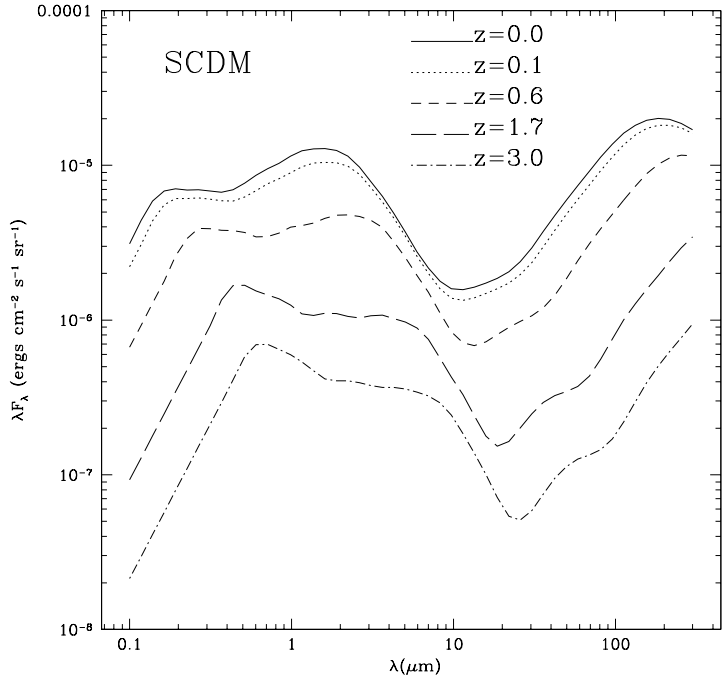,height=10truecm,width=10truecm}}
\caption{The extra-galactic background light from the far UV to the far IR, for
the SCDM model. The dotted and dashed lines show the contribution to the $z=0$
EBL from galaxies at $z\geq 0.1$, 0.6, 1.7, and 3.0 as indicated. Most of the
observed EBL at the wavelengths shown is produced by relatively nearby
galaxies. }
\label{fig:ebl}
\end{figure}
The semi-analytic models provide all the necessary information to compute the
integrated extra-galactic background light. For this purpose, we have
implemented a simple model for dust extinction in which the mass of dust in
each galaxy is assumed to be proportional to the mass of metals in the cold
gas. We then require that the energy absorbed by dust must be re-radiated
according to an assumed dust emission spectrum. The dust models are summarized
in \cite{mythesis} and are effectively very similar to those used by
Ref.~\cite{guiderdoni}, except that we have not attempted to include the
heavily extinguished ultra-luminous starburst galaxies which probably
contribute significantly to the far IR background \cite{guiderdoni}. This will
require more detailed modelling of dust absorption and re-radiation in
starburst galaxies, which we will pursue in future studies. Fig.~\ref{fig:ebl}
shows the resulting integrated extra-galactic background light (EBL) for the
fiducial SCDM models of Ref.~\cite{sp98}. The contribution to the present day
EBL from galaxies at various redshifts is also shown. Apparently, most of the
observed EBL is produced by relatively nearby galaxies. This is mainly due to
the usual decrease of energy density with the expansion of the universe. The
EBL from 0.1 to 2 microns is mainly starlight. At $\lambda \la 1 \mu$, a
significant portion of the light is absorbed by dust and reradiated in the
far-IR. The EBL at $\ga 10 \mu$ is almost completely due to re-radiation by
dust, and is therefore quite sensitive to our assumed dust absorption and
emission models. The characteristic shape of the EBL is also sensitive to the
assumed cosmology, star formation history, and IMF \cite{bullock}. With the
wealth of multi-band observations that are now becoming available, the EBL will
be an important probe of galaxy formation and cosmology.

\subsection{Summary}
We have shown that our CDM-based hierarchical models are in good agreement with
direct observational estimates of the global star formation history, and are
compatible with the fossil evidence which indicates that a significant fraction
of the stars in galaxies today formed at high redshift. An important new
ingredient in our models is the inclusion of starbursts triggered by
galaxy-galaxy interactions. It follows naturally in the CDM picture of
structure formation that such starburst events are more common at high
redshift, and are also brighter because the gas supply has not yet been
depleted. The starburst picture accounts naturally for many of the observed
features of the high redshift Lyman-break galaxies, including their number
density and luminosity distribution, and also their sizes, star formation
rates, colors, ages, and line-widths. In addition, the starburst mode results
in a significant modification of the star formation history of the universe,
moving the peak in the Madau diagram to higher redshift and leading to a much
more gradual decline in the star formation rate out to $z\sim6$. This is in
better agreement with the ``new Madau diagram'' that results when the
considerable dust extinction at $z\ga3$ now favored by observers is taken into
account, and with the SCUBA observations.

Many of the galaxy mergers that trigger starbursts will result in the formation
of an early-type galaxy or bulge. Small mass starburst remnants may be accreted
by larger galaxies, forming a Pop II halo \cite{trager} like that of our own
Galaxy \cite{kvj}. At high redshift ($z\ga2$), merger events occur mainly in
and around massive, strongly clustered dark matter halos. The merger-starburst
picture therefore predicts simultaneously the strong observed clustering of the
Lyman-break galaxies \cite{kolatt:mergers}, and that the oldest stars in the
universe should be found preferentially in dense environments today. The mass
of old stars ($z_{\rm form} > 3$) predicted by the starburst models is in good
agreement with observational estimates, unlike some previous CDM-based models,
which did not produce enough old stars. The gas density and metal content of
the universe as a function of redshift, indicated by observations of damped
Lyman-$\alpha$ systems, are important complementary constraints on the star
formation history and on recipes used to model star formation and supernovae
feedback. We present a calculation of the integrated extra-galactic background
light, which will be an increasingly important constraint on galaxy formation
and the associated processes of dust absorption and re-radiation.

In recent years, our knowledge of the Universe has been considerably extended
both in redshift and in wavelength, and no doubt this trend will continue. The
interpretation of these exciting observations is subject to many uncertainties
due to poorly understood astrophysical processes, notably star formation, dust
extinction, and the stellar initial mass function. We argue that the best
approach to overcoming these problems is to investigate many different kinds of
constraints. We conclude that the CDM framework, as realized by the
complementary techniques of simulations and semi-analytic models of galaxy
formation, can account for the current observations and provides an excellent
basis for continuing study of these issues.

\section*{Acknowledgments}
We thank our collaborators George Blumenthal, James Bullock, Avishai
Dekel, S.M. Faber, Michael Gross, Patrik Jonsson, Anatoly Klypin,
Tsafrir Kolatt, Andrey Kravtsov, Yair Sigad, Ian Walker, and Risa
Wechsler for allowing us to report on our new work here. JRP
acknowledges support from NASA and NSF grants at UCSC, and a Forchheimer
Visiting Professorship at Hebrew University. RSS acknowledges support
from a NASA grant and a GAANN Fellowship at UCSC and a University
Fellowship from the Hebrew University, Jerusalem.

\vfill
\end{document}